**Stretched Exponential in Non-Linear Stochastic Field Theories**

Moshe Schwartz

School of Physics and Astronomy

Tel Aviv University

Ramat Aviv, Tel Aviv 69978,

Israel

and

S. F. Edwards

Cavendish Laboratory, Madingley

Road CB3OHE Cambridge

United Kingdom

Abstract

We consider the time dependent two point function, $<\phi_q(t)\phi_{-q}(0)>$ in non-linear stochastic field theories, for which the KPZ equation serves as a prototype, in particular we consider small q's and long times such that $\omega_q t \gg 1$ ($\omega_q$ being the corresponding decay rate). We find that, since the generic case has $\omega_q \propto q^\mu$ for small q where $\mu > 1$, the decay of the two point function is given by a stretched exponential in $\omega_q t$ multiplied by a power of t, $<\phi_{-q}(0)\phi_q(t)> \propto t^{\beta_d} \exp[-\gamma(\omega_q t)^{1/\mu}]$, where $\beta_d = (d-1)/2\mu$, d is the dimensionality of space and $\gamma$ a dimensionless constant.

It is well known that many relaxation processes in condensed matter systems are characterized by stretched exponential decay. The list of examples contains relaxation in glasses[1-3] and in polymer systems[4,5], dielectric and viscoelastic relaxation[6,7] etc. The theoretical derivation of stretched exponential decay involves usually extensive numerical simulations. This is true not only in cases where a realistic model is chosen to describe the physical system[8] but also in cases where more simplified models are used[10-13].

The present letter has a number of goals. The first is to show that phenomena of stretched exponential decay are much more abundant than can be expected from the canonical list presented above. Many interesting macroscopic systems such as magnets at the critical points, a turbulent fluid or a surface evolving under ballistic deposition etc., are non linear and share the property, that a disturbance of wave vector $\vec{q}$ decays with a characteristic decay rate $\omega_q$, that scales at small $q$'s as some power of $q$, $\omega_q \propto q^\mu$. (Clearly, such systems attracted a lot of interest over a long period of time but that interest has not focused on the question of long time decay, that we address here). In most cases $\mu > 1$. An important exception is the case of turbulence where $\mu$ is about 2/3. We will show that for such systems that are generically described in terms of appropriate stochastic non-linear field theories and have $\mu > 1$, the decay is characterized by a stretched exponential function. The special case of turbulence will be discussed in the future. Our second goal is to construct a simple analytic framework within which the stretched exponential decay is derived. Our third goal is to show that the derivation is generic and that when the long time behaviour is considered most of the detail of the specific theory is not relevant. The only relevant parameter will be shown to be the exponent $\mu$ characterizing the small $q$ dependence of the $q$ dependent decay rate $\omega_q \propto q^\mu$.

We consider a scalar field $h(\vec{r},t)$ with Fourier transform $h_q(t)$ obeying a non-linear stochastic equation of the form

$$\frac{\partial h_q}{\partial t} + \nu_q h_q + \sum M_{q\ell m} h_\ell h_m + \sum N_{q\ell mn} h_\ell h_m h_n + .. = \eta_q, \qquad (1)$$

where M,N have $\ell + m = q, \ell + m + n = q$ etc. and $\eta_q(t)$ is a noise term obeying

$$<\eta_q> = 0 \quad \text{and} \quad <\eta_q(t)\eta_{-q}(t')> = 2D^o(q)\delta(t-t') \qquad (2)$$

The KPZ equation, for example, has N=0 and $M = \delta_{\ell+m-q} \vec{\ell} \cdot \vec{m} \frac{g}{\sqrt{\Omega}}$, where $\Omega$ is the volume of the system (to be taken eventually to infinity) and g is the non-linear coupling constant. For the dynamical $\phi^4$ theory M=0 and $N = \delta_{\ell+m+n-q} \frac{g}{\Omega}$ (with the appropriate upper cut-off restriction) etc. The generalization of the above to vector fields needed in turbulence is straightforward. In order to be specific we will consider in the following the case where only M is different from zero but the derivation will readily suggest the treatment appropriate in other cases.

In our previous work[14,15] we considered such systems and obtained the steady state "structure factor", $\phi_q = <h_q h_{-q}>_s \propto q^{-\Gamma}$ for small $q$ and the corresponding decay rate, $\omega_q$, given by

$$\omega_q^{-1} = \frac{\int_o^\infty <h_q(t)h_{-q}(0)> dt}{<h_q h_{-q}>_s} \propto q^{-\mu} \text{ for small } q, \quad (3)$$

The method we employed is a self consistent expansion around a linear model that produces the "exact" $\phi_q$ and $\omega_q$.

The expansion has thus the form[14-16]

$$\phi_q = \phi_q + c_q\{\phi_p, \omega_p\} \quad , \quad (4)$$

explicitly, in second order of the expansion,

$$\phi_q = \phi_q + \frac{1}{\omega_q}\{D_o(q) - v_q \phi_q + \sum_{\ell,m} \frac{M^2_{q\ell m} \phi_\ell \phi_m}{\omega_q + \omega_\ell + \omega_m} + 2\sum \frac{M_{q\ell m} M_{\ell q m}}{\omega_q + \omega_\ell + \omega_m} \phi_q \phi_\ell\} \quad , \quad (5)$$

and a similar relation for $\omega$

$$\omega_q = \omega_q + d_q\{\phi_p, \omega_p\} \quad . \quad (6)$$

The structure factor and decay rate are obtained by solving the coupled non-linear integral equations $c_q\{\phi_p, \omega_p\} = 0$ and $d_q\{\phi_p, \omega_p\} = 0$. In contrast to other expansions, the full correction, in a given order of the expansion, of the relevant physical quantities, is really small. In fact, it is <u>chosen</u> to be zero. In the following we obtain the "dynamical structure factor" $\Phi_q(t) = <h_q(0)h_{-q}(t)>_s$, using the same idea of a self consistent expansion. The average $<\text{---}>_s$ denotes that $h_q$ is measured in steady state at time $t = 0$ and then $h_{-q}$ is measured at some later time $t$. The "dynamical structure factor" $\Phi_q(t)$ normalized by $\phi_q = \Phi_q(0)$, the static structure factor, is thus a measure of the persistence in steady state of disturbances with wave vector $\vec{q}$.

We will proceed first to obtain an equation for $\Phi_q(t)$ and then to show that in the small $q$ and long time regime $(\omega_q t >> 1)$ it must show a stretched exponential decay. Note that as it will become obvious from the derivation, the long time decay depends on the specifics of the problem only through the dependence of $\mu$ $(\omega_q \propto q^\mu)$ on those specifics (as long as $\mu > 1$). We derive the equation for $\Phi_q(t)$ in detail but indeed the most important part is the final simplified equation (11) that holds in the longtime regime.

Our starting point is the field equation for $h_{q\omega}$, the Fourier transform in time $h_q(t)$.

$$i\omega h_{q\omega} + v_q h_{q\omega} + \sum C_{q\ell m} h_{\ell\sigma} h_{m\tau} = \eta_{q\omega} \quad , \tag{7}$$

where $C_{q\ell m} = \dfrac{M_{q\ell m}}{\sqrt{T}}$, T being an assumed periodicity in time to be taken eventually to infinity, $\sigma + \tau = \omega$ and the noise correlations are $<\eta_{q\omega}\eta_{-q-\omega}> = 2D^o(q)$. (Note, that in the d+1 dimensional space that includes time, the noise is quenched disorder !)

In the Chapman-Enskog spirit (as done in ref. 16) the equation is written in the form

$$\left[(i\omega + \omega_q)h_{q\omega} - \eta_{q\omega}^o\right] + \lambda\left[\sum_{\ell\sigma,m\tau} C_{q\ell m} h_{\ell\sigma} h_{m\tau} - \eta_{q\omega}^{(1)}\right] + \lambda^2\left[(v_q - \omega_q)h_{q\omega}\right] = 0 \quad , \tag{8}$$

where $\lambda$ is going to be taken as 1 but is used at present as an indicator to show the construction of the perturbation expansion as an expansion in $\lambda$. The noise is split into two terms $\eta_{q\omega}^{(o)}$ and $\eta_{q\omega}^{(1)}$ such that $<\eta_{q\omega}^{(o)}\eta_{-q-\omega}^{(o)}> = D_{q\omega}$ and the correct $\Phi_{q\omega}$ is given by $\Phi_{q\omega} = \dfrac{D_{q\omega}}{\omega_q^2 + \omega^2}$. This choice implies that ignoring the $\lambda$ and $\lambda^2$ terms in eq. (8), we still obtain from a linear equation the correct $\Phi_{q\omega}$. Eq. (8) enables to obtain $h_{q\omega}$ explicitly to second order in $\lambda$. The expression for $h_{q\omega}$ is multiplied into its complex conjugate and only terms up to second order in $\lambda$ are retained and then $\lambda$ is set to be 1 and the expression is averaged over $\eta_{q\omega}$. The result is an equation of the form $\Phi_{q\omega} = \Phi_{q\omega} + e\{\Phi_{p\sigma}\}$. Equating $e\{\Phi_{p\sigma}\}$ to zero yields

$$\left[\omega^2 + v_q^2\right]\Phi_{q\omega} - 4\sum_{\ell\sigma m\tau} C_{q\ell m}C_{\ell qm}^* \left\{\frac{[-i\omega + \omega_q]}{[i\sigma + \omega_\ell]} + \frac{[i\omega + \omega_q]}{[-i\sigma + \omega_\ell]}\right\}\Phi_{m\tau}\Phi_{q\omega}$$
$$-2\sum_{\ell\sigma m\tau} |C_{q\ell m}|^2 \Phi_{\ell\sigma}\Phi_{m\tau} = D^o(q) \tag{9}$$

Defining $C_{q\ell m} = \dfrac{A_{\ell,q-\ell}\delta_{\ell+mq}}{\sqrt{\Omega}}$, letting $\Omega$ and T tend to infinity and Fourier transforming in $\omega$, we obtain

$$-\left[\frac{\partial^2}{\partial t^2} - v_q^2\right]\Phi_q(t) - \frac{4}{(2\pi)^{d-1}}\int d^d\ell A_{\ell,q-\ell}A^*_{q,\ell-q} \times \{\int_o^\infty dt' e^{-\omega_\ell t'}\Phi_{q-\ell}(t')$$
$$\left[\frac{\partial}{\partial t}\Phi_q(t-t') + \omega_q\Phi_q(t-t')\right] + \int_{-\infty}^o dt' e^{\omega_\ell t'}\Phi_{q-\ell}(t')\left[-\frac{\partial}{\partial t}\Phi_q(t-t') + \omega_q\Phi_q(t-t')\right]\}$$
$$-\frac{2}{(2\pi)^{d-1}}\int d^d\ell |A_{\ell,q-\ell}|^2 \Phi_\ell(t)\Phi_{q-\ell}(t) = D^o(q)\delta(t) \qquad (10)$$

(Note that we define $\Phi_q(t) = \frac{1}{2\pi}\int \Phi_{q\omega}e^{i\omega t}d\omega$ and that $\Phi_q(t)$ is symmetric in time). The first step in the simplification of eq. (10) is to bring out $\Phi_q(t)$ of the integration over $t'$. If we assume that $\Phi_q(t)$ falls off as a function of $t$ slower than an exponential, it is possible to show that the most dominant contribution to the integral comes from the region with $t'$ very small compared to $t$. This enables a simplification that leads indeed to the result that $\Phi_q(t)$ decays slower than an exponential. Therefore the original assumption is consistent and the simplified equation reads

$$\alpha_q\Phi_q(t) - \frac{2}{(2\pi)^{d-1}}\int d^d\ell |A_{\ell,q-\ell}|^2 \Phi_\ell(t)\Phi_{q-\ell}(t) = 0 , \qquad (11)$$

where

$$\alpha_q = v_q^2 - \frac{8\omega_q}{(2\pi)^{d-1}}\int d^d\ell A_{\ell,q-\ell}A_{q,\ell-q}\int_o^\infty e^{-\omega_\ell t'}\Phi_{q-\ell}(t')dt' \qquad (12)$$

and where we have neglected $\frac{\partial^2}{\partial t^2}\Phi_q(t)$ compared to $\Phi_q(t)$, that is also consistent with $\Phi_q(t)$ decaying slower than an exponential.

It is important to note here that there are many ways of obtaining equations for $\Phi_q(t)$ and indeed using different approximation schemes we obtained equations that differ from eq. (10). The important point however, is that their simplified form is similar to eq. (11) and leads to the same form of decay.

The long time form of the solution is expected to be given by

$$\Phi_q(t) = \psi_q^{(1)}f_1(\omega_q t) + \psi_q^{(2)}f_2(\omega_q t) + ... , \qquad (13)$$

where $\lim_{x \to \infty} \frac{f_{n+1}(x)}{f_n(x)} = 0$.

(Therefore, there is no reason to assume apriori that $\psi_q^{(1)} = \phi_q$. In fact, if the series described by eq. (14) converges for all times $\phi_q = \sum_i \psi_q^{(i)}$ if the $f_i$'s are chosen to have $f_i(0) = 1$).

The leading order form, $f_1$, will be obtained now by balancing in eq. (11) the terms that are most dominant at long times. It is most important that apart from assuming that the appropriate integrals converge, the following discussion does not depend at all on the specific form of the $M_{q\ell m}$ (or equivalently $A_{\ell,q-\ell}$).

Consider next a simple algebraic fact. The function $\Delta_\theta(\vec{\ell},\vec{q}) = \ell^\theta + |\vec{q}-\vec{\ell}|^\theta - q^\theta$ is negative in a small region of $\vec{\ell}$ (for fixed $\vec{q}$) for $\theta > 1$. (For $\theta < 1$ $\Delta_\theta(\vec{\ell},\vec{q})$ is always positive). This implies that due to the fact that $\mu > 1$, $f_1$ cannot decay as an exponential or faster. If that were the case, the last term on the left hand side of eq. (11) would have a contribution from that small region of $\vec{\ell}$ discussed above, that cannot be balanced against other terms in the equation, because it decays slower.

To obtain the leading decay consider first the one dimensional case. The only possible way, by which the leading behaviour of the two terms left in equation (11) can be balanced is by having $f_1(x) = \exp[-\gamma x^{1/\mu}]$. The reason is that, say, for positive q there is a finite segment $0 \leq \ell \leq q$ on which $\omega_q^{1/\mu} = \omega_{q-\ell}^{1/\mu} + \omega_\ell^{1/\mu}$. (It is obvious that $\omega_q \propto q^\mu$ only for small q but on the segment $0 \leq \ell \leq q$ small q implies small $\ell$ as well as small $q - \ell$, therefore, $\omega_\ell \propto \ell^\mu$ and $\omega_{q-\ell} \propto (q-\ell)^\mu$).

The balancing condition leaves us with an equation relating the coefficients $\psi_q^{(1)}$

$$\alpha_q \psi_q^{(1)} = \frac{2}{(2\pi)^{d-1}} \int_o^q d\ell \left|A_{\ell,q-\ell}\right|^2 \psi_\ell^{(1)} \psi_{q-\ell}^{(1)}. \tag{14}$$

The full solution involves the calculation of the coefficients $\psi_q^{(1)}$ from the equation above. The important point, however, is the stretched exponential form obtained for $f_1$. The $\ell > q$ and $\ell < 0$ regions of integration produce a correction term of the form $\exp[-\gamma q t^{1/\mu}]$ times a negative power of t. The correction terms have to be balanced against correction terms to the time dependent structure factor (eq. (13)).

It is immediately clear that a stretched exponential cannot solve eq. (11) for dimensions larger than 1. If we take $f(x) = \exp[-\gamma x^\delta]$ with $\delta > 1/\mu$ the $\Phi_q(t)$ term will dominate. If $\delta < 1/\mu$ the integral term will dominate, so that we are forced again at having $\delta = 1/\mu$. The problem is that the relation $[\omega_{\vec{q}}]^{1/\mu} = [\omega_{\vec{\ell}}]^{1/\mu} + [\omega_{\vec{q}-\vec{\ell}}]^{1/\mu}$ holds only for vectors $\vec{\ell}$ of size bounded from above by $q$ and parallel to vector $\vec{q}$ i.e., on a set of measure zero. Therefore, even for $\delta = 1/\mu$, $\Phi_q(t)$ dominates over the integral. The solution must, therefore, be of the form

$$f_1(x) = x^\beta \exp[-\gamma x^{1/\mu}] \quad \text{with} \quad \beta > 0. \tag{15}$$

To determine $\beta$, we have to consider the effective region of $\vec{\ell}$ integration in eq. (11). Let, $\vec{\ell} = \vec{\ell}_{II} + \vec{\ell}_\perp$, where $\vec{\ell}_{II}$ is the part parallel to $\vec{q}$ and $\vec{\ell}_\perp$ the part perpendicular to it. The region of integration is of order q,t independent, in the direction of $\vec{\ell}_{II}$. The integration in the $\vec{\ell}_\perp$ directions is cut-off effectively at $\frac{\ell_\perp^2}{q} \propto t^{-1/\mu}$. This implies that the exponent $\beta$ is given by

$$\beta = \frac{d-1}{2\mu} \ . \tag{16}$$

The derivation presented above focuses on the case where the non linear part of the stochastic field equation has in it only a product of two fields (like in the KPZ equation). When more than two fields are involved (like in the case of critical dynamics in the $\phi^4$ theory) the situation is the same. The first term on the left hand side of eq.(11) is linear $\Phi_q(t)$ and the second term contains an integral on $\ell_1, \ell_2, ..., \ell_k$ and the product $\Phi_{\ell_1}(t)\Phi_{\ell_2}(t)...\Phi_{\ell_k}(t)\Phi_{q-\sum \ell_i}(t)$. The algebraic fact mentioned above is true for any number of variables. Namely, for a set of vector $\vec{\ell}_i$ obeying $\sum_{i=1}^N \vec{\ell}_i = \vec{q}$,

$$\sum_{i=1}^N |\vec{\ell}_i|^\theta - q^\theta > 0 \quad \text{for} \quad \theta < 1 \tag{17}$$

and for $\theta > 1$ there is a small region of the $\ell's$, for fixed $\vec{q}$, where this expression becomes negative. Therefore, the derivation of the stretched exponential behaviour follows in the same way as above. The power $\beta$ is also unaffected. It is easy to see using the arguments given above for the case of two fields in the non-linear part, that for $n$ fields the equation determining $\beta$ is

$$\beta = n\beta - \frac{d-1}{2\mu}(n-1) . \tag{18}$$

The long time form of $\Phi_q(t)$ is given therefore by

$$\Phi_q(t) = c\phi_q [\gamma q t^{1/\mu}]^{\frac{d-1}{2}} \exp[-\gamma q t^{1/\mu}], \qquad (19)$$

where $c$ is a dimensionless constant not necessarily 1!. The above form seems to be robust against various changes in the self consistent expansion and even against inclusion of higher orders. This will be discussed in future publications.